\documentclass[aoas,MSNbibl,nameyear,seceqn,rotating,dvips]{arximspdf}
\usepackage{dcolumn}
\usepackage{graphicx}

\doi{10.1214/13-AOAS630} 
\volume{7}
\issue{3}
\pubyear{2013}
\firstpage{1386}
\lastpage{1420}

\makeatletter

\newcolumntype{d}[1]{D{.}{.}{#1}}

\newcommand{\N}{\mathrm{N}}

\newcommand{\sd}{\operatorname{sd}}
\newcommand{\Var}{\operatorname{Var}}

\newcommand{\E}{\mathrm{E}}
\newcommand{\bX}{\mathbf{X}}
\newcommand{\bW}{\mathbf{W}}
\newcommand{\bQ}{\mathbf{Q}}

\makeatother

\begin{document}
\begin{frontmatter}

\title{Assessing lack of common support in causal inference using
Bayesian nonparametrics:
Implications for evaluating the effect of breastfeeding on
children's cognitive outcomes\thanksref{T2}}
\runtitle{Assessing lack of common support in causal inference}

\thankstext{T2}{Supported in part by the Institute of Education
Sciences Grant R305D110037
and by the Wang Xuelian Foundation.}

\begin{aug}
\author[A]{\fnms{Jennifer} \snm{Hill}\corref{}\ead[label=e1]{jennifer.hill@nyu.edu}}
\and
\author[B]{\fnms{Yu-Sung} \snm{Su}\ead[label=e2]{suyusung@tsinghua.edu.cn}}
\runauthor{J. Hill and Y.-S. Su}
\affiliation{New York University and Tsinghua University}
\address[A]{New York University\\
Steinhardt, HMSS\\
246 Greene St. 3rd Floor\\
New York, New York 10003\\
USA\\
\printead{e1}}
\address[B]{Department of Political Science\\
Tsinghua University\\
Haidian District, Beijing, 100084\\
China\\
\printead{e2}} 
\end{aug}

\received{\smonth{8} \syear{2011}}
\revised{\smonth{1} \syear{2013}}

%
\begin{abstract}
Causal inference in observational studies typically requires making
comparisons between groups that are dissimilar. For instance,
researchers investigating the role of a prolonged duration of
breastfeeding on child outcomes may be forced to make comparisons
between women with substantially different characteristics on average.
In the extreme there may exist neighborhoods of the covariate space
where there are not sufficient numbers of both groups of women (those
who breastfed for prolonged periods and those who did not) to make
inferences about those women. This is referred to as lack of common
support. Problems can arise when we try to estimate causal effects for
units that lack common support, thus we may want to avoid inference
for such units. If ignorability is satisfied with respect to a set of
potential confounders, then identifying whether, or for which units,
the common support assumption holds is an empirical question. However,
in the high-dimensional covariate space often required to satisfy
ignorability such identification may not be trivial. Existing methods
used to address this problem often require reliance on parametric
assumptions and most, if not all, ignore the information embedded in
the response variable. We distinguish between the concepts of ``common
support'' and ``common causal support.'' We propose a new approach for
identifying common causal support that addresses some of the
shortcomings of existing methods. We motivate and illustrate the
approach using data from the National Longitudinal Survey of Youth to
estimate the effect of breastfeeding at least nine months on reading
and math achievement scores at age five or six. We also evaluate the
comparative performance of this method in hypothetical examples and
simulations where the true treatment effect is known.
\end{abstract}

%

\end{frontmatter}

\section{Introduction}\label{secintro}
Causal inference strategies in observational studies that assume
ignorability of the treatment assignment also typically require an
assumption of common support; that is, for binary treatment assignment,
$Z$, and a vector of confounding covariates, $\mathbf{X}$, it is commonly
assumed that $0 < \Pr(Z=1 \mid\mathbf{X}) < 1$. Failure to satisfy this
assumption can lead to unresolvable imbalance for matching methods,
unstable weights in inverse-probability-of-treatment
weighting (IPTW) estimators, and undue reliance on model specification in
methods that model the response surface.

To satisfy the common support assumption in practice, researchers have
used various strategies to identify (and excise) observations
in neighborhoods of the covariate space where there exist only
treatment units (no controls) or only control units (no treated)
[see, e.g., \citet{heckichitodd1997}].
Unfortunately many of these methods rely on correct
specification of a model for the treatment assignment. Moreover, all
such strategies (that we have identified) fail to take advantage of
the outcome variable, $Y$, which can provide critical information
about the relative importance of each potential confounder. In
the extreme this information could help us discriminate between situations
where overlap is lacking for a variable that is a true confounder versus
situations when it is lacking for a variable that is not predictive of the
outcome (and thus not a true confounder). Moreover, there is currently
a lack of guidance regarding
how the researcher can or should characterize how
the inferential sample has changed after units have been discarded.

In this paper we propose a strategy to address the problem of
identifying units that lack common support, even in fairly
high-dimensional space. We start by defining the causal inference
setting and estimands of interest ignoring the common support issue.
We then review a causal inference strategy
[discussed previously in \citet{hill2011}] that exploits an
algorithm called Bayesian Additive Regression Trees
[BART; Chipman, George and McCulloch (\citeyear{chipgeormccu2007,chipgeormccu2010})].
We discuss the issue of common support and then introduce the
concept of ``common causal support.''

Our method for addressing
common support problems exploits a key feature of the BART approach
to causal inference. When BART is used to estimate causal effects one
of the ``byproducts'' is that it yields individual-specific posterior
distributions for each potential outcome; these act as proxies for the
amount of information we have about these outcomes. Comparisons of
posterior distributions of counterfactual outcomes versus factual (observed)
outcomes can be used to create red flags when the amount of information about
the counterfactual outcome for a given observation is not sufficient to warrant
making inferences about that observation. We illustrate this method in several
simple hypothetical examples and examine the performance of our strategy
relative to propensity-based methods in simulations. Finally, we
demonstrate the practical differences in our breastfeeding example.

\section{Causal inference and BART}\label{secCasualInferenceAndBart}
This section describes notation, estimands, and assumptions
followed by a discussion of how BART can be used to estimate causal
effects.\setcounter{footnote}{1}\footnote{Green and Kern
(\citeyear{greekern2012}) discuss extensions to this BART
strategy for causal inference to more thoroughly
explore heterogeneous treatment effects.}

\subsection{Notation, estimands and assumptions}
We discuss a situation where we attempt to identify a causal effect
using a sample of independent observations of size $n$. Data for the
$i$th observation consists of an outcome variable, $Y_i$, a~vector of
covariates, $\bX_i$, and a binary treatment assignment variable,
$Z_i$, where $Z_i=1$ denotes that the treatment was received.
We define potential outcomes for this observation,
$Y_i(Z_i=0)=Y_i(0)$ and $Y_i(Z=1)=Y_i(1)$, as the
outcomes that would manifest under each of the treatment
assignments. It follows that $Y_i = Y_i(0)(1-Z_i) + Y_i(1)Z_i$.
Given that observational samples are rarely random samples from the
population and we will be limiting our samples in further nonrandom
ways in order to address lack of overlap, it makes sense to focus on
sample estimands
such as the conditional average treatment effect (CATE),
$\sum_{i=1}^n E[Y_i(1) - Y_i(0) \mid X_i]$, and the conditional
average treatment
effect for the treated (CATT), $\sum_{i:Z_i=1} E[Y_i(1) - Y_i(0) \mid X_i]$.
Other common sample estimands we may consider are the
sample average treatment effect (SATE), $\sum_{i=1}^n E[Y_i(1) - Y_i(0)]$,
and the sample average effect of the treatment on the treated (SATT),
$\sum_{i:Z_i=1} E[Y_i(1) - Y_i(0)]$.

If ignorability holds for our sample, that is, $Y_i(0), Y_i(1) \perp
Z_i \mid\bX_i=\mathbf{x}$, then $\E[Y_i(0) \mid\bX_i=\mathbf{x}] =
\E[Y_i \mid Z_i=0, \bX_i=\mathbf{x}]$ and $\E[Y_i(1)
\mid\bX_i=\mathbf{x}] = \E[Y_i \mid Z_i=1$, $\bX_i=\mathbf{x}]$. The
basic idea behind the BART approach to causal inference is to assume
$\E[Y_i(0) \mid\bX=\mathbf{x}]=f(0,\mathbf{x})$ and $\E[Y_i(1)
\mid\bX_i=\mathbf{x}]=f(1,\mathbf{x})$ and then fit a very
flexible model for $f$.

In principle, any method that flexibly estimates $f$ could be used to
model these conditional expectations.
Chipman, George and McCulloch (\citeyear{chipgeormccu2007,chipgeormccu2010}) describe BARTs advantages
as a predictive algorithm compared to similar alternatives in the data mining
literature. \citet{hill2011} describes the
advantages of using BART for causal inference estimation over several
alternatives common in the causal inference literature.

The BART algorithm consists of two pieces: a sum-of-trees model and a
regularization prior. Dropping the $i$ subscript for notational convenience,
we describe the sum-of-trees model by $Y = f(z, \mathbf{x}) +
\varepsilon
$, where
$\varepsilon\sim N(0,\sigma^2)$ and
\[
f(z, \mathbf{x}) = g(z,\mathbf{x};T_1,M_1) + g(z,
\mathbf{x};T_2,M_2) + \cdots+ g(z,\mathbf{x};T_m,M_m).
\]
Here each $(T_j,M_j)$ denotes a single subtree model. The number of
trees is typically allowed to be large [Chipman, George and McCulloch
(\citeyear{chipgeormccu2007,chipgeormccu2010}) suggest 200, though, in
practice, this number should not exceed the number of observations in
the sample]. As is the case with related sum-of-trees strategies (such
as boosting), the algorithm requires a strategy to avoid overfitting.
With BART this is achieved through a regularization prior that allows
each $(T_j,M_j)$ tree to contribute only a small part to the overall
fit.

BART fits the sum-of-trees model using a MCMC algorithm that cycles
between draws of $(T_j,M_j)$ conditional on $\sigma$ and draws of
$\sigma$ conditional on all of the $(T_j,M_j)$. Converence can be
monitored by plotting the residual standard deviation parameter
$\sigma$ over time. More details regarding BART can be found in
Chipman, George and McCulloch
(\citeyear{chipgeormccu2007,chipgeormccu2010}).

It is straightforward to use BART to
estimate average causal effects such as $E[Y(1) \mid X=x]- E[Y(0) \mid
X=x] =
f(1,x) - f(0,x)$. Each iteration of the BART Markov Chain generates a
new draw of $f$ from the posterior distribution. Let $f^r$ denote the
$r$th draw of $f$. To perform causal inference, we then compute
$d_i^r=f^r(1,\mathbf{x}_i) - f^r(0,\mathbf{x}_i)$, for $i = 1,\ldots, n$.
If we average the $d_i^r$
values over $i$ with $r$ fixed, the resulting values will be our Monte
Carlo approximation to the posterior distribution of the average
treatment effect for the associated population. For example, we
average over the entire sample if we want to estimate the average
treatment effect. We average over ${i\dvtx z_i=1}$ if we want to
estimate the effect of the treatment on the treated.

\subsection{Past evidence regarding BART performance}\label{ssecPastEvidence}
Hill (\citeyear{hill2011}) provides evidence of superior performance of
BART relative
to popular
causal inference strategies in the context of nonlinear response
surfaces. The focus
in those comparisons is on methods that are reasonably simple to understand
and implement: standard linear regression, propensity score matching
(with regression
adjustment), and inverse probability of treatment weighted linear regression
[IPTW; \citet{imbe2004,kurtetal2006}].

One vulnerability of BART identified in \citet{hill2011} is that
there is nothing
to prevent it from extrapolating over areas of the covariate space where
common support does not exist. This problem is not unique to BART; it
is shared by all causal modeling strategies that do not first discard
(or severely downweight) units in these areas. Such extrapolations
can lead to biased inferences because of the lack of information
available to identify either $\E[Y(0) \mid X]$ or $\E[Y(1) \mid X]$ in
these regions. This paper proposes strategies to address this issue.

\subsection{Illustrative example with one predictor}\label
{ssecExampleOnePredictor}
We illustrate use of BART for causal inference with an
example [similar to one used in \citet{hill2011}]. This example
also demonstrates
both the problems that can occur when common support is compromised and
a potential solution.

%
\begin{figure}

\includegraphics{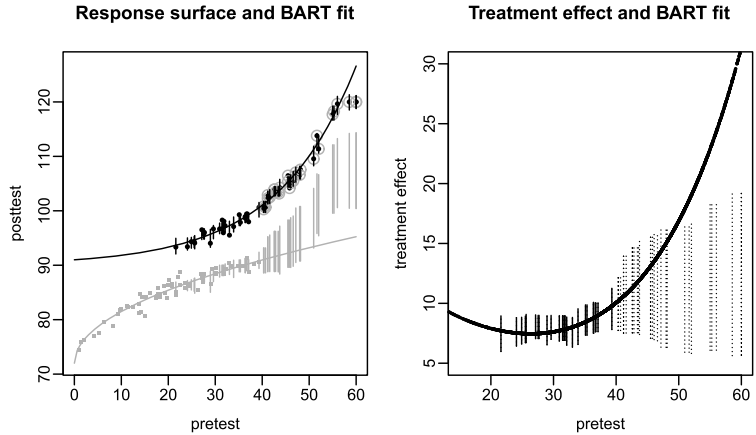}

\caption{Left panel: simulated data (points) and true response
surfaces. The black upper curve and
points that follow it correspond to the treatment condition; the grey
lower curve and points
that follow it correspond to the control condition. BART inference for
each treated
observation is displayed as a 95\% posterior interval for $f(1,x_i)$
and $f(0,x_i)$.
Discarded units (described in Section \protect\ref{secexamples}) are circled.
Right panel: solid curve represents the treatment effect as it varies
with our pretest, $X$.
BART inference is displayed as 95\% posterior intervals for the
treatment effect
for each treated unit. Intervals for discarded units (described in Section
\protect\ref{secexamples}) are displayed as dotted lines. In this sample
the conditional average treatment effect for the treated (CATT) is 12.2,
and the sample average treatment effect for the treated (SATT) is 11.8.}
\label{bartfitsimple}
\end{figure}

Figure \ref{bartfitsimple} displays simulated data from each of two
treatment groups from a hypothetical educational intervention. The
120 observations were generated independently
as follows. We generate the treatment variable as $Z \sim\operatorname
{Bernoulli}(0.5)$.
We generate a pretest measure as $X \mid Z=1 \sim\N(40,10^2)$ and
$X \mid Z=0 \sim\N(20,10^2)$. Our post-test potential outcomes are
drawn as $Y(0)\mid X \sim\N(72 +3\sqrt{X}, 1)$ and
$Y(1)\mid X \sim\N(90 + \exp(0.06X),1)$. Since we conceptualize
both our confounder and our outcome as test scores, a ceiling is imposed
on each (60 and 120, resp.). Even with this constraint this is an
extreme example of heterogeneous treatment effects, designed, along
with the lack of overlap, to make it
extremely difficult for any method to successfully estimate the true
treatment effect.

In the left panel, the upper solid black curve represents $E[Y(1) \mid
X]$ and the lower grey one $E[Y(0) \mid X]$. The black circles close to
the upper curve are the treated and the grey squares close to the lower
curve are the untreated (ignore the circled points for now). Since
there is only one confounding covariate, $X$, the difference between
the two response surfaces at any level of $X$ represents the treatment
effect for observations with that value of the pretest $X$. In this
sample the conditional average treatment effect for the treated (CATT)
is 12.2, and the sample average treatment effect for the treated (SATT)
is 11.8.

A linear regression fit to the data yields a substantial underestimate, 7.1
(s.e. 0.62), of both estimands.
Propensity score matching (not restricted to common support) with
subsequent regression adjustment yields a much better estimate,
10.4~(s.e. 0.52), while the IPTW regression estimate
is 9.6 (s.e. 0.45). For both of these methods the propensity
scores were estimated using logistic regression.

The left panel of Figure \ref{bartfitsimple} also displays the BART
fit to the response
surface (with number of trees equal to 100 since there are only 120
observations).
Each vertical line segment corresponds to individual level inference
about either
$E[Y_i(0) \mid X_i]$ or $E[Y_i(1) \mid X_i]$ for each treated observation.
Note that the fit is quite good until we try to predict $E[Y_i(0) \mid X_i]$
beyond the support of the data. The right panel displays the true
treatment effect as
it varies with $X$, $E[Y(1)-Y(0) \mid X]$, as a solid curve. The BART inference
(95\% posterior interval) for the treatment effect for each treated
unit is superimposed
as a vertical segment (ignore the solid versus dashed distinction for now).
These individual-level inferences can be averaged to obtain
inference for the effect of the treatment on the treated which is 9.5
with 95\%
posterior interval (7.7, 11.8); this interval best corresponds to
inference with
respect to the conditional average treatment effect on the treated
[\citet{hill2011}].

None of these methods yields a 95\% interval that captures CATT.
BART is the only method to capture SATT, though at the expense of
a wider uncertainty interval. All the approaches are hampered by the
fairly severe lack of common support. Notice, however, the way that the
BART-generated
uncertainty bounds grow much wider in the range where there is no
overlap across treatment groups ($X > 40$). The marginal intervals
nicely cover the true conditional treatment effects until we start to
leave this neighborhood. However, inference in this region is
based on extrapolation. Our goal is to devise a rule to determine how
much ``excess'' uncertainty should warrant removing a unit from the
analysis. We will return to this example in Section~\ref{secexamples}.

\section{Identifying areas of common support}\label{secIdentifyCommonSupport}
It is typical in causal inference to assume common support. In
particular, many researchers assume ``strong ignorability''
[\citet{roserubi1983}]
which combines the standard ignorability assumption discussed
above with an assumption of common support often formalized as
$0 < \Pr(Z \mid\bX) < 1$. It is somewhat less common for researchers
to check whether common support appears to be empirically satisfied for
their particular data set.

Moreover, the definition of common support is itself left
vague in practice. Typically, $\bX$ comprises the set of covariates the
researcher has chosen to justify the ignorabilty
assumption. As such, conservative researchers will understandably include
a large number of pretreatment variables in $\bX$.
However, this will likely mean that $\bX$
includes any number of variables that are not required to satisfy
ignorability once we condition on some other subset of the vector
of covariates. Importantly, the requirement of common support need
not hold for the variables not in this subset, thus, trying to force common
support on these extraneous variables can lead to
unnecessarily discarding observations.

The goal instead should be to ensure \textit{common causal support}
which can be defined as $0 < \Pr(Z \mid\bW) < 1$, where
$\bW$ represents any subset of $\bX$ that will satisfy
$Y(0), Y(1) \perp Z \mid\bW$. Because BART takes advantage of
the information in the outcome variable, it should be better able to target
common causal support as will be demonstrated in the examples
below. Propensity score methods, on the other hand, ignore this
information, rendering them incapable of making these distinctions.

If the common causal support assumption does not hold for the units in our
inferential sample (the units in our sample about whom we'd
like to make causal inference), we do not have direct empirical evidence
about the counterfactual state for them. Therefore, if we retain these
units in our sample, we run the risk of obtaining biased treatment effect
estimates.

One approach to this problem is to weight observations by the strength
of support [for an example of this strategy in a propensity score
setting, see \citet{crumimbeetal2009}]. This strategy may yield
efficiency gains over simply discarding problematic units. However,
this approach has two key disadvantages. First, if there are a large
number of covariates, the weights may become unstable. Second, it
changes the interpretation of the estimand to something that may have
little policy or practical relevance. For instance, suppose the units
that have the most support are those currently receiving the program,
however, the policy-relevant question is what would happen to those
currently not receiving the program. In this case the estimand would
give most weight to those participants of least interest from a policy
perspective.

Another option is to identify and remove observations in neighborhoods
of the covariate space that lack sufficient common causal support.
Simply discarding observations deemed problematic is unlikely to lead
to an optimal solution. However, this approach has the advantage of greater
simplicity and transparency. More work will need to be done, however, to
provide strategies for adequately profiling the discarded observations
as well
as those that we retain for inference; this paper will provide a simple starting
point in this effort. The primary goal of this paper is simply to
describe a
strategy to identify these problematic observations.

\subsection{Identifying areas of common causal support with BART}
\label{ssecIdentifyCommonCausalSuppportBART}
The simple idea is to capitalize on the fact that the posterior
standard deviations of the individual-level conditional expectations estimated
using BART increase markedly in areas that lack common causal support, as
illustrated in Figure~\ref{bartfitsimple}.
The challenge is to determine how extreme these standard deviations should
be before we need be concerned. We present several possible rules for
discarding units. In all strategies when implementing BART
we recommend setting the ``number of trees'' parameter to 100
to allow BART to better determine
the relative importance of the variables.

Recall that the individual-level causal effect for each unit can
be expressed as $d_i=f(1,\mathbf{x}_i) - f(0,\mathbf{x}_i)$.
For each unit, $i$, we have explicit information about
$f(Z_i,\mathbf{x}_i)$.
Our concern is whether we have enough information about $f(1-Z_i,
\mathbf{x}_i)$.
The amount of information is reflected in the posterior standard deviations.
Therefore, we can create a metric for assessing our uncertainty\vspace*{1pt} regarding
the sufficiency of the common support for any given unit by comparing
$\sigma_i^{f_0} = \sd(f(0,\mathbf{x}_i))$ and $\sigma
_i^{f_1} = \sd(f(1,\mathbf{x}_i))$,
where $\sd(\cdot)$ denotes the posterior standard deviation.
In practice, of course we use Monte Carlo approximations to these quantities,
$s_i^{f_0}$~and~$s_i^{f_1}$, respectively, obtained by calculating the
standard deviation of the draws of $f(0,\mathbf{x}_i)$ and
$f(0,\mathbf{x}_i)$
for the $i$th observation.

\subsubsection*{BART discarding rules} Our goal is to use the
information that
BART provides to create a rule for determining which units lack sufficient
counterfactual evidence (i.e., residing in a neighborhood
without common support). For example, when estimating the effect of
the treatment on units, $i$, for which $Z_i=a$, one might consider
discarding any unit, $i$, with $Z_i=a$, for which $s_i^{f_{1-a}} > m_a$,
where $m_a = \max_j \{ s_j^{f_a} \}$, $\forall j\dvtx Z_j=a$.
So, for instance,\vspace*{1pt} when estimating the effect of the treatment on the treated
we would discard treated units whose counterfactual standard deviation
$s_i^{f_0}$ exceeded the maximum standard deviation under the
\textit{observed} treatment condition $s_i^{f_1}$ across all the
treated units.

This cutoff is likely too sharp, however, as even chance disturbances might
put some units beyond this threshold. Therefore, a more useful rule
might use
a cutoff that includes a ``buffer'' such that we would only discard for
unit $i$
in the inferential group defined as those with $Z_i=a$, if
\[
s_i^{f_{1-a}} > m_a + \sd\bigl(
s_j^{f_a}\bigr)\qquad \mbox{(\textit{1 sd rule})},
\]
where $\sd(s_j^{f_a})$ represents the estimated standard deviation of the
empirical distribution of $s_j^{f_a}$ over all units with $Z_j=a$.
For this rule to be most useful, we need
$\Var(Y \mid\bX, Z=0) = \Var(Y \mid\bX, Z=1)$ to hold at least
approximately.

Another option is to consider the squared ratio of posterior standard
deviations (or,
equivalently, the ratio of posterior variances) for each observation,
with the counterfactual
posterior standard deviation in the numerator. An approximate
benchmark distribution for this ratio might be a $\chi^2$ distribution
with 1 degree
of freedom. Thus, for an observation with $Z_i=a$ we can choose cutoffs that
correspond to\vadjust{\goodbreak} a specified $p$-value of rejecting the hypothesis that
the variances are
the same of 0.10,
\[
\bigl(s_i^{f_{1-a}}/s_i^{f_a}
\bigr)^2 > 2.706\qquad\forall i\dvtx z_i=1 \mbox{ (\textit{$
\alpha=0.10$ rule})}
\]
or a $p$-value of 0.05,
\[
\bigl(s_i^{f_{1-a}}/s_i^{f_a}
\bigr)^2 > 3.841\qquad\forall i\dvtx z_i=1 \mbox{ (\textit{$
\alpha=0.05$ rule})}.
\]
These ratio rules do not require the same type of homogeneity of
variance assumption across
units as does the \textit{1 sd rule}. However, they rest instead on an
implicit assumption of homogeneity of variance within unit across
treatment conditions. Additionally, they may
be less stable and will be prone to rejection for units that have
particularly large amounts of information for the observed state. For
instance, an observation in a neighborhood of the covariate space that
has control units may still reject (i.e., be flagged as a discard)
if there
are, relatively speaking, many more treated units in this neighborhood
as well.

\subsubsection*{Exploratory analyses using measures of common causal
support uncertainty}
Another way to make use of the information in the posterior standard
deviations is
more exploratory. The idea here is to use a classification strategy
such as a regression
tree to identify neighborhoods of the covariate space with relatively
high levels of
common support uncertainty. For instance, when the goal is estimation
of the
effect of the treatment on the treated we may want to determine
neighborhoods that
have clusters of units with relatively high levels of $s_i^{f_{1-Z_i}}$ or
$s_i^{f_{1-Z_i}}/s_i^{f_{Z_i}}$. Then these ``flags'' combined with
researcher knowledge of the substantive context of the research problem can
be combined to identify observations or neighborhoods to be excised from
the analysis if it is deemed necessary. This approach may have the
advantage of
being more closely tied to the science of the question being addressed.
We illustrate possibilities for exploring and characterizing these neighborhoods
in Sections \ref{ssecprofiling} and \ref{secRealExample}.

Reliance on this type of exploratory strategy will likely be eschewed
by researchers who
favor strict analysis protocols as a means of promoting honesty in research.
In fact, the original BART causal analysis strategy was conceived with this
predilection in mind, which is why (absent the need or desire to address
common support issues) the advice given is to run it only once and at
the default
settings; this minimizes the amount of researcher ``interference''
[\citet{hill2011}].
These preferences may still be satisfied, however, by specifying one of
the discarding
rules above as part of the analysis protocol. For further discussion of this
issue see Section \ref{sechonesty}.

\subsection{Competing strategies for identifying common support}
\label{secCompetingStrategies}
The primary competitors to our strategy for identification of units
that lack
sufficient common causal support rely on propensity scores. While\vadjust{\goodbreak}
there is little advice directly given to the topic of how to use the propensity
score to identify observations that lack common support for the included
predictors [for a notable exception see \citet{crumimbeetal2009}],
in practice, most researchers using propensity score strategies
first estimate the propensity score and then discard any inferential
units that extend beyond the range of the propensity score
[\citet{heckichitodd1997,dehewahb1999,morghard2006}].
This type of exclusion is performed automatically in at least two popular
propensity score matching software packages, MatchIt in R [\citet
{hoimaikingstua2011}]
and psmatch2 in Stata [\citet{leuvsian2011}] when the ``common support''
option is chosen.
For instance, if the focus is on the effect of the treatment on the
treated, one would typically discard the treated units with propensity scores
greater than the maximum control propensity score, unless there happened
to be some treated with propensity scores less than the minimum control
propensity score (in which case these treated units would be discarded as
well).

More complicated caliper matching methods might further discard inferential
units that lie within the range of propensity scores of their
comparison group if
such units are more than a set distance (in propensity score units)
away from
their closest match [see, e.g., \citet{frol2004}]. Given the
number of
different radius/caliper matching methods and the lack of clarity about
the optimal
caliper width, it was beyond the scope of this paper to examine those
strategies as well.

Weighting methods are typically not coupled with discarding rules since one
of the advantages touted by weighting advocates is that IPTW allows the
researcher
to include their full sample of inferential and comparison units.
However, in some
situations failure to discard inferential units that are quite
different from the bulk
of the comparison units can lead to more unstable weight estimates.

We have two primary concerns about use of propensity scores to identify
units that
fail to satisfy common causal support. First, they require a correct
specification of
the propensity score model. Offsetting this concern is the fact that
our BART strategy
requires a reasonably good fit to the response surface. As demonstrated
in \citet{hill2011},
however, BART appears to be flexible enough to perform well in this
respect even with
highly nonlinear and nonparallel response surfaces. A further caveat to
this concern
is the fact that several flexible estimation strategies have recently
been proposed for
estimating the propensity score. In particular, Generalized Boosted Models
(GBM) and Generalized Additive Models (GAM) have both been advocated
in this capacity with mostly positive results
[\citet{mccaridgmorr2004,wooreitkarr2008}],
although some more mixed findings exist for GBM in particular settings
[\citet{hillweisszhai2011}].
In Section~\ref{secsimulation} we explore the
relative performance of these approaches against our BART approach.

Our second concern is that the propensity score strategies ignore the
information
about common support embedded in the response variable. This can be important
because the researcher typically never knows which of the covariates
in her data set are actually confounders; if a covariate is not
associated with both the treatment assignment and the outcome, we need
not worry about forcing overlap with regard to it. Using propensity
scores to
determine common support gives greatest weight to those variables that
are most predictive of the treatment variable. However, these
variables may not be most important for predicting the outcome. In
fact, there is no guarantee that they are predictive of the outcome
variable at all. Conversely, the propensity score may give insufficient
weight to variables that are highly predictive of the outcome and thus
may underestimate the risk of retaining units with questionable
support with regard to such a variable.

The BART approach, on the other hand, naturally and coherently
incorporates all of this information. For instance, if there is lack
of common support with respect to a variable that is not strongly
predictive of the outcome, then the posterior standard deviation for
the counterfactual unit should not be systematically higher to a large
degree. However, a variable that similarly lacks common support but
is strongly predictive of the outcome should yield strong differences
in the distributions of the posterior standard deviations across
counterfactuals. Simply put, the standard deviations should pick up
``important'' departures from complete overlap and should largely
ignore ``unimportant'' departures. This ability of BART to capitalize
on information in the outcome variable allows
it to more naturally target \textit{common causal support}.

\subsection{Honesty}\label{sechonesty}
Advocates of propensity score strategies sometimes directly advocate for
ignoring the information in the response variable [\citet{rubi2002}].
The argument goes that such practice allows the
researcher to be more honest because a propensity score model can in
theory be chosen (through balance checks) before the outcome variable
is even included in the analysis. This approach can avoid the potential
problem of repeatedly tweaking a model until the treatment effect
meets one's prior expectations. However, in reality there is nothing to
stop a researcher from estimating a treatment effect every time he
fits a new propensity score model and, in practice, this surely happens.
We argue that a better way to achieve this type of honesty is to fit just
one model and use a prespecified discarding rule,
as can be achieved in the BART approach to causal inference.

\section{Illustrative examples}\label{secexamples}
We illustrate some of the key properties of our method using several
simple examples. Each example represents just one draw from the
given data generating mechanism, thus, these examples are not meant
to provide conclusive evidence regarding relative performance of the
methods in each scenario. These examples provide an\vadjust{\goodbreak} opportunity to visualize
some of the basic properties of the BART strategy relative to more
traditional propensity score strategies: propensity score matching
with regression adjustment and IPTW regression estimates. Since
we estimate average treatment effects for the treated in all the examples,
for the IPTW approach the treated units all receive weights of 1 and the
control units receive weights of $\hat{e}(x)/(1-\hat{e}(x))$, where
$\hat{e}(x)$ denotes the estimated propensity score.

\subsection{Simple example with one predictor}\label
{ssecexampleWithOnePredictor}
First, we return to the simple example from Section \ref
{secCasualInferenceAndBart} to see how our common
causal support identification strategies work in that setting. Since
there is only
one predictor and it is a true confounder, common support and common causal
support are equivalent in this example and we would not expect to see much
difference between the methods.

The circled treated observations in the left-hand panel of Figure \ref
{bartfitsimple}
indicate the 29 observations that would be dropped using the standard propensity
score discard rule. Similarly, the dotted line segments in the right
panel of the figure
indicate individual-specific treatment effects that would no longer be
included in our
average treatment
effect inference. All three BART discard rules lead to the same set of discarded
observations as the propensity score strategy in this example.

SATT and CATT for the remaining units are 7.9 and 8.0, respectively.
Our new BART
estimate is 8.2 with 95\% posterior interval (7.7, 9.0). With this
reduced sample
propensity score, matching (with subsequent regression adjustment)
yields an
estimate of the treatment effect at 8.3 (s.e. 0.26) while IPTW yields
an estimate of 7.6 (s.e. 0.32).

Advantages of BART over the propensity score approach are not evident
in this
simple example. They should manifest in examples where the assignment
mechanism is more difficult to model or when there are multiple potential
confounders and not all variables that predict treatment also predict
the outcome
(or they do so with different emphasis). We explore these issues next.

\subsection{Illustrative examples with two predictors}\label
{ssecexampleWithTwoPredictors}
We now describe two\break slightly more complicated examples to illustrate the
potential advantages of BART over propensity-score-based competitors.
In both examples there are two independent covariates, each generated
as $\N(0,1)$, and the goal is to estimate CATT which is equal to 1 (in fact,
the treatment effect is constant across observations in these examples).
The question in each case is whether some of the treated observations should
be dropped due to lack of empirical counterfactuals.

\subsubsection{Example \textup{2A}: Two predictors, no confounders}
\label{sssectwoPredictorsNoConfounders}
\begin{figure}

\includegraphics{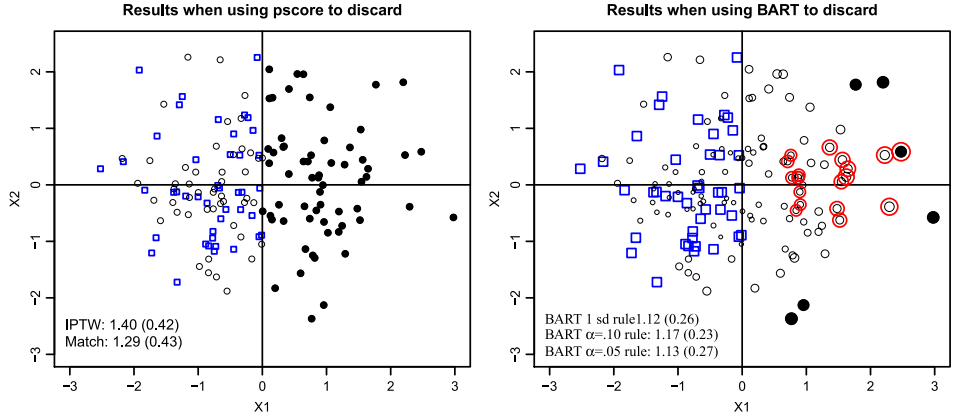}

\caption{Plots of simulated data with two predictors; the true treatment
effect is 1. $X_1$ predicts
treatment assignment only and $X_2$ predicts outcome only. Control
observations are displayed as squares. Treated observations
are displayed as circles. The left panel displays results
based on propensity score common support; solid
circles indicate which observations were discarded. In the right panel
the size of the circle is proportional to the $s_i^{f_0}$. Observations
discarded based on the BART \textup{1 sd rule} are displayed as solid circles.
Observations discarded based
on the BART \textup{$\alpha=0.10$ rule} are circled. No observations were
discarded based on the BART \textup{$\alpha=0.05$ rule} ratio rule.}
\label{exmp1}
\end{figure}

In the first example the assignment mechanism is simple---after\vadjust{\goodbreak}
generating $Z$ as a random flip of the coin, all controls with $X_1>0$
are removed. The response surface is generated as $E[Y \mid Z,X_1,X_2] =
Z + X_2 + X_2^2$, thus, the true treatment effect is constant at 1.
Since there are no true confounders in this example,
the requirement of common support on both $X_1$ and $X_2$
will be overly conservative; overlap on neither is required to
satisfy common causal support. Figure \ref{exmp1} illustrates how each
strategy performs in this scenario.

%

In both plots circles represent treated observations and squares represent
control observations. The left panel shows the results based on discarding
units that lack common support with respect to the propensity score.
The observations discarded by the propensity score method are displayed
as solid circles. Since treatment
assignment is driven solely by $X_1$, there is a close mapping between
$X_1$ and the propensity score (were it not for the fact that $X_2$
was also in the estimation model for the propensity score, the
correspondence would be one-to-one). 62 of the 112 treatment observations
are dropped based on lack of overlap with regard to the propensity
score.\looseness=1

After re-estimating the propensity score matching on the smaller sample,
the matching estimate is 1.29. Since treatment assignment is independent
of the potential outcomes by design, this estimate should be unbiased over
repeated samples. However, it now has less than half the observations
available for estimation. Inverse-probability-of-treatment weighting (IPTW)
yields an estimate of 1.40 (s.e.~0.42) after
discarding.\footnote{If we fail to re-estimate the propensity score after
the initial discard, the matching estimate is 1.53 (s.e. 0.40) and the
IPTW estimate is 1.47 (s.e. 0.44).}

In the right plot of Figure \ref{exmp1} the size of the
circle for each treated unit is proportional to the corresponding size of
the posterior standard deviation of the expected outcome under the
control condition (in this case, the counterfactual condition for the
treated). The size of the square that represents each control
observation is proportional to the cutoff level for discarding units.
Observations discarded by the \textit{1~sd rule} have been made
solid. Observations discarded by the \textit{$\alpha=0.10$ rule} have
been circled.
No observations were discarded using the \textit{$\alpha=0.05$ rule}.

In contrast to the propensity score discard rule, the BART \textit{1 sd rule}
recognizes that $X_1$ does not play an important role in the response
surface, so it only drops 7 observations that are at the boundary of the
covariate space. The corresponding BART estimate is 1.12 with a posterior
standard deviation (0.26) that is quite a bit smaller than the standard errors
of both propensity score strategies. The \textit{$\alpha=0.10$ rule}
drops 18 observations,
on the other hand, and these observations are in a different neighborhood
than those dropped by the \textit{1 sd rule} since the individual level ratios
can get large not just when $s_i^{f_0}$ is (relatively) large but also when
$s_i^{f_1}$ is (relatively) small. The corresponding estimate of 1.17 and
associated standard error (0.23) are quite similar to those achieved by the
\textit{1 sd rule}. The BART \textit{$\alpha=0.05$ rule} yields an
estimate from the
full sample since it leads to no discards (1.13 with a standard error
of 0.27).
All of the BART strategies benefit from being able to take advantage
of the information in the outcome variable.

\subsubsection{Example \textup{2B}: Two predictors, changing information}
\label{sssectwopredschanginginfo}
In the second example the assignment mechanism is slightly more complicated.
We start by generating $Z$ as a binomial draw with probabilities
equal to the
inverse logit of $X_1 + X_2 - 0.5X_1X_2$. Next all control units with $X_1>0$
\textit{and} $X_2>0$ are removed. Two different response surfaces
are generated, each as $\E[Y \mid Z,X_1,X_2] = Z + 0.5X_1 + 2X_2 + \phi
X_1 X_2$,
where one version sets $\phi$ to 1 and the other sets $\phi$ to 3.
Therefore, both covariates are confounders in this example and both the common
support assumption and the common causal support assumption are in question.
Once again the treatment effect is 1.

%
\begin{figure}

\includegraphics{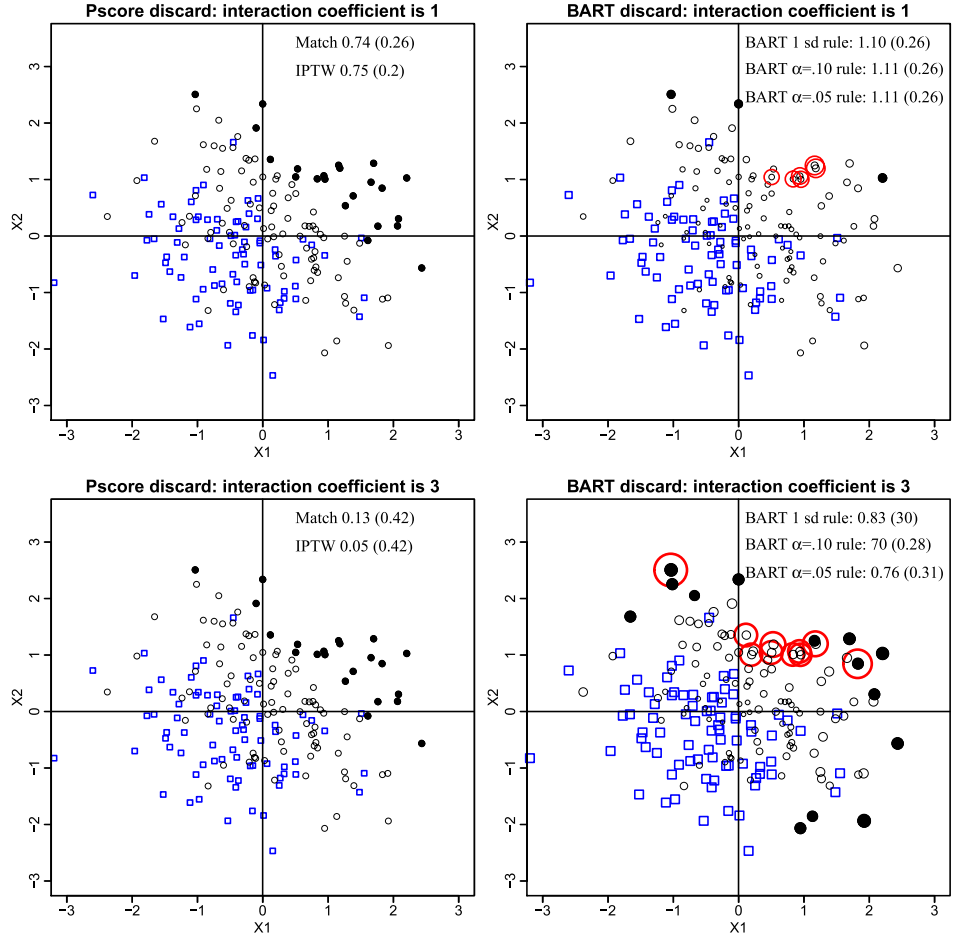}

\caption{Plots of simulated data with two predictors; the true treatment
effect is 1. The display is analogous to Figure \protect\ref{exmp1},
although here
the two left plots display propensity score results across the two scenarios
and the two right display BART results across the two scenarios.}
\label{exmp2}
\end{figure}

The propensity score discard strategy chooses the same
observations to discard across both response surface scenarios because
it only takes into account information in the assignment mechanism.
Thus, the left panel in Figure \ref{exmp2} presents the same plot twice;
the only differences are the estimates of the treatment effect which
vary with response surface. The matching estimates get worse (0.74, then
0.13) as the response surface becomes more highly nonlinear
as do the IPTW estimates (0.75, then 0.05). The uncertainty associated
with the estimates grows between the first and second response surface
(from roughly 0.2 to roughly 0.4), yet standard 95\% confidence
intervals do
not cover the truth in the second setting.\footnote{If we fail to re-estimate
the propensity score after discarding, the estimates are just as bad or worse.
For the first scenario, the matching estimate would be 0.65 (s.e. 0.28) and
the IPTW estimate would be 0.75 (s.e. 0.20). For the second scenario, the
matching estimate would be 0.02 (s.e. 0.44) and the IPTW estimate would
be 0.06 (s.e.~0.36).\looseness=1}

The BART discard strategies, on the other hand, respond to information
in the response surface. Since the lack of overlap occurs in an area
defined by the intersection of $X_1$ and $X_2$, uncertainty in the
posterior counterfactual predictions increases sharply when the coefficient
on the interaction moves from 1 to 3 (as displayed in the top and bottom
plots in the right panel of Figure \ref{exmp2}, resp.) and more
observations are dropped for both the \textit{1 sd rule} and \textit
{$\alpha=0.10$ rule}. In this
example \textit{$\alpha=0.10$ rule} once again focuses more on
observations in the quadrant
with lack of overlap with respect to the treatment condition, whereas
\textit{1 sd rule}
identifies observations than tend to have greater uncertainty more generally.
No observations are dropped by \textit{$\alpha=0.05$ rule} even when
$\phi$ is 3.

The BART treatment effect estimates in both the first scenario (all
about 1.1)
and the second scenario (0.83, 0.70 and 0.76) are all closer to the truth
than the
propensity-score-based estimates in this example. In the first scenario the
uncertainty estimates (posterior standard errors of 0.26 for each) are slightly
higher than the standard errors for the propensity score estimates;
in the second scenario the uncertainty estimates (posterior standard
errors all around 0.3)
are all smaller than the standard errors for the propensity score estimates.

\subsection{Profiling the discarded units: Finding a needle in a
haystack}\label{ssecprofiling}

When treatment effects are not homogeneous, discarding observations
from the inferential group can change the target estimand.
For instance, if focus is on the effect of the treatment on the treated
(e.g., CATT or SATT) and we discard treated
observations, then we can only make inferences about the treated units
that remain (or the population they represent). It is
important to have a sense of how this new estimand differs from the
original. In this section we illustrate a simple way to ``profile'' the
units that remain in the inferential sample versus those that were
discarded in an attempt to achieve common support.

In this example there are 600 observations and 40 predictors, all
generated as
$\N(1,1)$. Treatment was assigned randomly at the outset; control observations
were then eliminated from two neighborhoods in this high-dimensional
covariate space. The first such neighborhood is defined by $X_3 > 1$
and $X_4 > 1$,
the second by \mbox{$X_5 > 1$} and $X_6 > 1$. The nonlinear
nonparallel response surface is generated as
$E[Y(0) \mid\bX] <- 0.5X_1 + 2X_2 + 0.5X_5 + 2X_6 + X_5X_6 + 0.5X_5^2 +
1.5X_6^2$ and
$E[Y(1) \mid\bX] <- 0.5X_1 + 2X_2 + 0.5X_5 + 2X_6 + 0.2X_5X_6 $.
The treatment effect thus varies across levels of the included covariates.
Importantly, since $X_3$ and $X_4$ do not enter into the response
surface, only
the second of the two neighborhoods that lack overlap should be of concern.

\begin{figure}

\includegraphics{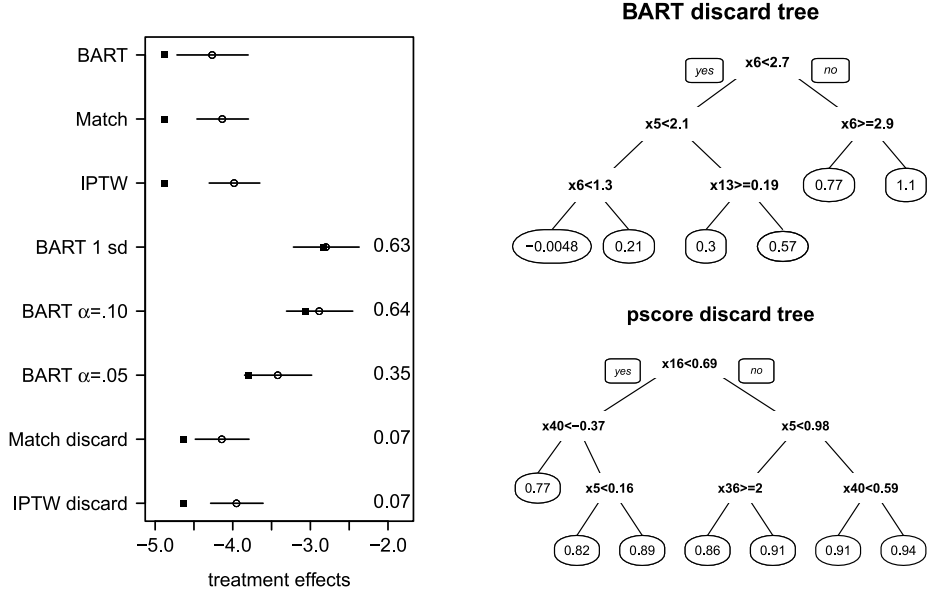}

\caption{Left plot displays estimands (squares) and attempted inference
(circles for estimates and bars for 95\% intervals) for the
BART and propensity score methods both with and without discarding.
The right plots display regression tree fits using the covariates as predictors.
The responses used are the statistic from \textup{1 sd rule} and then the
propensity score, respectively.}
\label{figprofiling}
\end{figure}

The leftmost plot in Figure \ref{figprofiling} displays results from
the BART and
propensity score methods both before and after discarding. The numbers
at the
right represent the percentage of the treated observations that were
dropped for\vadjust{\goodbreak}
each discard method. Solid squares represent the true estimand (SATT)
for the
sample corresponding to that estimate (the same for all methods that
do not discard
but different for those that do). Circles and line segments represent
estimates and corresponding 95\% intervals for each estimate. None of
the methods that fail to
discard has a 95\% interval that covers the truth for the full sample.
After discarding using the BART rules, all of the intervals cover the
true treatment
effect for the remaining sample. The propensity score methods drop far fewer
treated observations, leading to estimands that do not change much and estimates
that still do not cover the estimands for the remaining sample.

%

We make use of simple regression trees [CART; \citet
{brei1984,brei2001}] to
investigate the differences between the neighborhoods perceived as problematic
for each method. Regression trees use predictors to partition the
sample into
subsamples that are relatively homogenous with respect to the response variable.
For our purposes, the predictors are our potential confounders and the response
is the statistic corresponding to a given discard rule.\footnote{Another
strategy would be to use the indicator for discard as the response
variable. This
could become problematic if the number of discarded observations is
small and
would yield no information about the likelihood of being discarded in situations
where no units exceeded the threshold.}
A simple tree fit provides a crude means of describing the neighborhoods
of the covariate space considered most problematic by each rule with respect
to common support. Each tree is restricted to a maximum depth of three for
the sake of parsimony.

To profile the units that the BART \textit{1 sd rule} considers
problematic, we use
for the response variable in the tree the corresponding statistic
relative to the cutoff rule
(appropriate for estimating the effect of the treatment on the treated),
$s_i^{f_0} - m_1 - \sd( s_j^{f_1}) $, where $i$
and $j$ index treated units. The tree fit is displayed in the top right
plot of
Figure \ref{figprofiling} with the mean of the response in each
terminal node
given in the corresponding oval. Note that the decision rules for
the tree are based almost exclusively on the variables $X_5$ and $X_6$, as
we would hope they would be given how the data were generated.

The tree fit using the propensity score as the response is displayed in the
lower right plot of Figure \ref{figprofiling}. $X_5$ plays a far less
prominent
role in this tree and $X_6$ does not appear at all. $X_{16}$, $X_{36}$,
and $X_{40}$ play important roles even though these variables are not
strong predictors in the response surface; in fact, these are all independent
of both the treatment and the response.

This example illustrates two things. First, regression trees may be a useful
strategy for profiling which neighborhoods each method has identified as
problematic with regard to common support. Second, the propensity score
approach may fail to appropriately discover areas that lack overlap if the
model for the assignment mechanism and the model for the response
surface are not well aligned with respect to the relative importance
of each variable. We explore the importance of this type of alignment in
more detail in the next section.

\section{Simulation evidence}\label{secsimulation}
This section explores simulation evidence regarding the performance of our
proposed method for identifying lack of common support relative to the
performance of two commonly-used and several less-commonly-used
propensity-score-based alternatives. Overall we compare the performance
of 12 different estimation strategies across 32 different simulated scenarios.

\subsection{Simulation scenarios}\label{ssecSimulationScenarios}
These scenarios represent all combinations of five design factors. The
first factor varies whether the logit of the conditional expectation of
the treatment assignment is linear or nonlinear in the covariates. The
second factor varies the relative importance of the covariates with
regard to the assignment mechanism versus the response surface. In one
setting of this factor (``aligned'') there is substantial alignment in
the predictive strength of the covariates across these two
mechanisms---the covariates that best predict the treatment also
predict the outcome well. In the other setting (``not as aligned'') the
covariates that best predict the treatment strongly and those that
predict the response strongly are less well aligned (for details see
the description of the treatment assignment mechanisms and response
surfaces and Table \ref{figcoefalign}, below).\footnote{For a related
discussion of the importance of alignment in causal inference see Kern
et al. (\citeyear{kernetal2013}).} The third factor is the ratio of treated to control
(4:1 or 1:4) units. The fourth factor is the number of predictors
available to the researcher (10 versus 50, although in both cases only
8 are relevant). The fifth and final factor is whether or not the
nonlinear response surfaces are parallel across treatment and control
groups; nonparallel response surfaces imply heterogeneous treatment
effects.

In all scenarios each covariate is generated independently from
$X_j \sim\mbox{N}(0, 1)$. These column vectors comprise the matrix
$\bX$.
The general form of the linear treatment assignment mechanism is
$Z \sim\operatorname{Binomial}(n, p)$ with $p = \operatorname
{logit}^{-1}(\omega+
\bX\gamma^L)$,
where the offset $\omega$ is specified to create the appropriate ratio of
treated to control units. The nonlinear form of this assignment mechanism
simply includes some nonlinear transformations of the covariates in
$\bX$, denoted
as $\bQ$ with corresponding coefficients $\gamma^{\mathrm{NL}}$.
The nonzero coefficients for the terms in these models are displayed in
Table \ref{figcoefalign}.

We simulate two distinct sets of response surfaces that differ in both
their level of
alignment with the assignment mechanism and whether they are parallel.
Both sets used are nonlinear in the covariates and each set is
generated generally as
\begin{eqnarray*}
\E\bigl[Y(0) \mid\bX\bigr] & = & \N\bigl( \bX\beta^{\mathrm{L}}_0
+ \bQ\beta^{\mathrm{NL}}_0, 1\bigr),
\\
\E\bigl[Y(1) \mid\bX\bigr] & = & \N\bigl( \bX\beta^{\mathrm{L}}_1
+ \bQ\beta^{\mathrm{NL}}_1 + \tau, 1\bigr),
\end{eqnarray*}
where $\beta^{\mathrm{L}}_{z}$ is a vector of coefficients for the
untransformed
versions of the predictors $\bX$ and $\beta^{\mathrm{NL}}_{z}$ is a
vector of
coefficients for the transformed versions of the predictors captured in
$\bQ$.
In the scenarios with parallel response surfaces, $\tau$ (the constant
treatment effect) is 4, $\beta^{\mathrm{L}}_{0}=\beta^{\mathrm
{L}}_{1}$, and $\beta^{\mathrm{NL}}_{0}=\beta^{\mathrm{NL}}_{1}$
and both use the coefficients from
$Y(0)$ in Table \ref{figcoefalign} (only nonzero coefficients displayed).
In the scenarios with responses surfaces are not parallel, $\tau=0$,
and the nonzero
coefficients in the $\beta^{\mathrm{L}}_{z}$ and $\beta^{\mathrm
{NL}}_{z}$ are displayed
in Table \ref{figcoefalign}.

%
\begin{sidewaystable}
\tabcolsep=0pt
\textwidth=\textheight
\tablewidth=\textwidth
\caption{Nonzero coefficients in $\gamma^{\mathrm{L}}$ and $\gamma
^{\mathrm{L}}$ for
the treatment assignment mechanism as well as for $\beta^{\mathrm
{L}}_z$ and
$\beta^{\mathrm{NL}}_z$ for the nonlinear, not parallel response surfaces.
Coefficients for the parallel response surface are the same as those
for $Y(0)$ in the nonparallel response surface}
\label{figcoefalign}
\begin{tabular*}{\tablewidth}{@{\extracolsep{\fill}}lccccccd{1.1}d{1.1}ccd{1.1}cccccccccc@{}}
\hline
& \multicolumn{1}{c}{$\bolds{x_1}$} & \multicolumn{1}{c}{$\bolds{x_2}$} & \multicolumn{1}{c}{$\bolds{x_1^2}$} & \multicolumn{1}{c}{$\bolds{x_2^2}$} & \multicolumn{1}{c}{$\bolds{x_2x_6}$} & \multicolumn{1}{c}{$\bolds{x_5}$} & \multicolumn{1}{c}{$\bolds{x_6}$} & \multicolumn{1}{c}{$\bolds{x_7}$}
& \multicolumn{1}{c}{$\bolds{x_8}$} & \multicolumn{1}{c}{$\bolds{x_9}$} & \multicolumn{1}{c}{$\bolds{x_{10}}$} & \multicolumn{1}{c}{$\bolds{x_5^2}$} & \multicolumn{1}{c}{$\bolds{x_6^2}$} & \multicolumn{1}{c}{$\bolds{x_5x_6}$} & \multicolumn{1}{c}{$\bolds{x_5x_6x_7}$}
& \multicolumn{1}{c}{$\bolds{x_7^2}$} & \multicolumn{1}{c}{$\bolds{x_7^3}$} & \multicolumn{1}{c}{$\bolds{x_8^2}$} & \multicolumn{1}{c}{$\bolds{x_7x_8}$} & \multicolumn{1}{c}{$\bolds{x_9^2}$} & \multicolumn{1}{c@{}}{$\bolds{x_9x_{10}}$}\\
\hline
\multicolumn{22}{@{}l}{Treatment assignment mechanisms}\\
\quad Linear & & & & & & 0.4 & 0.2 & 0.4 & 0.2 & 0.4 & 0.4 & & & & & & & & &
& \\
\quad Nonlinear & & & & & & 0.4 & 0.2 & 0.4 & 0.2 & 0.4 & 0.4 & 0.8 & 0.8 &
0.5 & 0.3
& 0.8 & 0.2 & 0.4 & 0.3 & 0.8 & 0.5\\
[6pt]
\multicolumn{22}{@{}l}{Response surfaces, nonlinear and not parallel}\\
\multicolumn{22}{l}{Aligned}\\
\quad$Y(0)$ & & & & & & 0.5 & & 2 & & 0.5 & 2 & 0.4 & 0.8 & & & 0.5 & & 0.5 &
& 0.5 &
0.7\\
\quad$Y(1)$ & & & & & & 0.5 & & 1 & 0.5 & & 0.8 & & & 0.3 & & & & & & & \\
[6pt]
\multicolumn{22}{@{}l}{Not as aligned}\\
\quad$Y(0)$ & 0.5 & 2\hphantom{.0} & 0.4 & 0.5 & 1 & 0.5 & 2 & & & & & 0.5 & 1.5 & 0.7 & & & & &
& & \\
\quad$Y(1)$ & 0.5 & 0.5 & & & & 0.5 & 2 & & & & & & & 0.3 & & & & & & & \\
\hline
\end{tabular*}
\end{sidewaystable}

Table \ref{figcoefalign} helps us understand the alignment in
predictor strength
between the assignment mechanism and response surfaces for each of the two
scenarios. The ``aligned'' version of the response surfaces places
weight on the
covariates most predictive of the assignment mechanism (both the linear
and nonlinear pieces). There is no reason to believe that this
alignment occurs in
real examples. Therefore, we explore a more realistic scenario where coefficient
strength is ``not as aligned.''

We replicate each of the 32 scenarios 200 times and in each simulation run
we implement each of 12 different\vadjust{\goodbreak} modeling strategies. For each the
goal is to estimate the conditional average effect of the treatment on the
subset of treated units that were not discarded.

\subsection{Estimation strategies compared}
\label{ssecEstimationStrategiesCompared}

We compare three basic causal inference strategies without
discarding---BART [implemented as described above and in \citet{hill2011}
except using 100 trees], propensity score matching, and IPTW---with
nine strategies that involve discarding.

The first three discarding approaches discard using the \textit{1 sd
rule}, the
\textit{$\alpha=0.10$ rule},
and the \textit{$\alpha=0.05$ rule} and each is coupled with a BART
analysis of the
causal effect
on the remaining sample.\footnote{We do not re-estimate BART after discarding
but simply limit our inference to MCMC results from the nondiscarded
observations.}
The remaining 6 approaches are combinations of 3 propensity score
discarding strategies and 2 analysis strategies. The 3 propensity score
discard strategies vary by the estimation strategy for the propensity
score model: standard logit, generalized boosted regression model
[recommended for propensity score estimation by \citet{mccaridgmorr2004}],
and generalized additive models
[recommended for propensity score estimation by \citet{wooreitkarr2008}].
The 2 analysis strategies (each conditional on a given propensity score
estimation model)
are one-to-one matching (followed by regression adjustment)
and inverse-probability of treatment weighting (in the context of a
linear regression
model). In all propensity score strategies the propensity score is
re-estimated after
the initial units are discarded. The $y$-axis labels of the results figures
indicate these 12 different combinations of strategies. All strategies
estimate the
effect of the treatment on the treated.

We implement these models in several packages in R [\citet
{RDevelopmentCoreTeam2009}].
We use the \texttt{bart()} function in the \texttt{BayesTree} package
[\citet{Chipman2009}]
to fit BART models. For each BART fit, we allow the maximum number of trees
in the sum to be 100 as described in
Section \ref{ssecIdentifyCommonCausalSuppportBART} above.
To ensure the convergence of the MCMC in BART without having to check
for each simulation
run, we are conservative and let the algorithm run for 3500 iterations
with the first 500
considered burn-in. To implement the GBM routine, we use the \texttt{gbm()}
function of the \texttt{gbm} package [\citet{Ridgeway2007}]. In an
attempt to optimize the
settings for esimating propensity scores, we adopt the suggestions of
[\citet{mccaridgmorr2004}, 409] for the tuning parameters of the
GBM: $100$ trees, a
maximum of 4 splits for each tree, a small shrinkage value of $0.0005$,
and a random sample
of 50\% of the data set to be use for each fit in each
iteration.\footnote{In response to a suggestion by a reviewer we also
implemented this method
using the \texttt{twang} package in \texttt{R} [\citet
{ridgetal2012}] using the settings suggested in the vignette
(n.trees${}={}$5000, interaction.depth${}={}$2, shrinkage${}={}$0.01). This did not
improve the GBM results.} We use the \texttt{gam()} function of the
\texttt{gam}
package [\citet{Hastie2009}] to implement the GAM routine.

%
\begin{figure}

\includegraphics{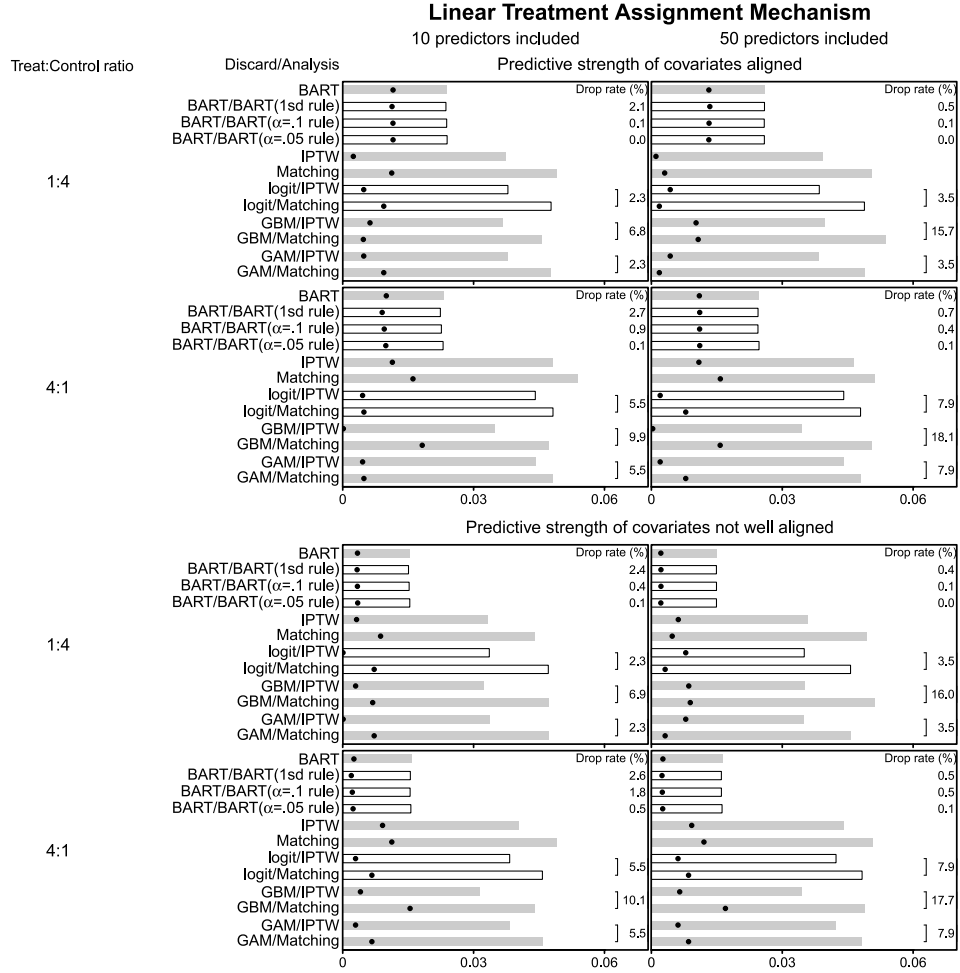}

\caption{Simulation results for the scenarios in which the treatment
assignment is linear and the response surfaces are parallel. Solid dots
represent average differences between estimated treatment effects and
the true ones standardized by the standard deviations of the outcomes.
Bars are root mean square errors (RMSE) of such estimates. The drop
rates are the percentage discarded units. Discard and analysis
strategies are described in the text. Five modeling strategies are
highlighted with hollow bars for comparison: the three BART strategies
and the most likely propensity scores versions to be implemented (these
are the same strategies illustrated in the examples in Section
\protect\ref{secexamples}).}\label{figsimltrpnl}
\end{figure}

\subsection{Simulation results}\label{simresults}

Figure \ref{figsimltrpnl} presents results from 8 scenarios that
have the common
elements of a \textit{linear treatment assignment mechanism} and
\textit{parallel response surfaces}. The linear treatment assignment mechanism
should favor the propensity score approaches. The top panel of 4 plots
in this figure
corresponds to the setting where there is alignment in the predictive
strength of the
covariates; this setting should favor the propensity score approach as
well since it implicitly
uses information about the predictive strength of the covariates with
regard to the
treatment assignment mechanism to gauge the importance of each
covariate as a
confounder. The bottom panel of Figure \ref{figsimltrpnl} reflects
scenarios in
which the predictive strength of the covariates is not as well aligned
between the
treatment assignment mechanism and the response surface. This setup provides
less of an advantage for the propensity score methods. The potential
for bias
across all methods, however, should be reduced.

Within each plot, each bar represents the root mean square error (RMSE)
of the
estimates for that scenario for a particular estimation strategy. The
dots represent
the absolute bias (the absolute value of the average difference
between the estimates and the CATT estimand). Drop rates for the discarding
methods are indicated on the right-hand side of each plot. We highlight
(with unfilled bars) the BART discard/analysis strategies as well as
the two propensity
score discard strategies that rely on the logit specification of the
propensity score
model (the most commonly used model for estimating propensity scores).

The first thing to note about Figure \ref{figsimltrpnl} is that
there is little bias
in any of the methods across all of these eight scenarios and likewise
the RMSEs
are all small. Within this we do see some small differences in the
absolute levels of bias across methods in the aligned scenarios, with slightly
less bias evidenced by the propensity score approaches and smaller RMSEs
for the BART approaches. In the nonaligned scenarios the differences in bias
nearly disappear (with a slight advantage overall for BART) and the
advantage with
regard to RMSE becomes slightly more pronounced. None of the methods drop
a large percentage of treated observations, but the BART rules discard
the least
(with one small exception).

%
\begin{figure}

\includegraphics{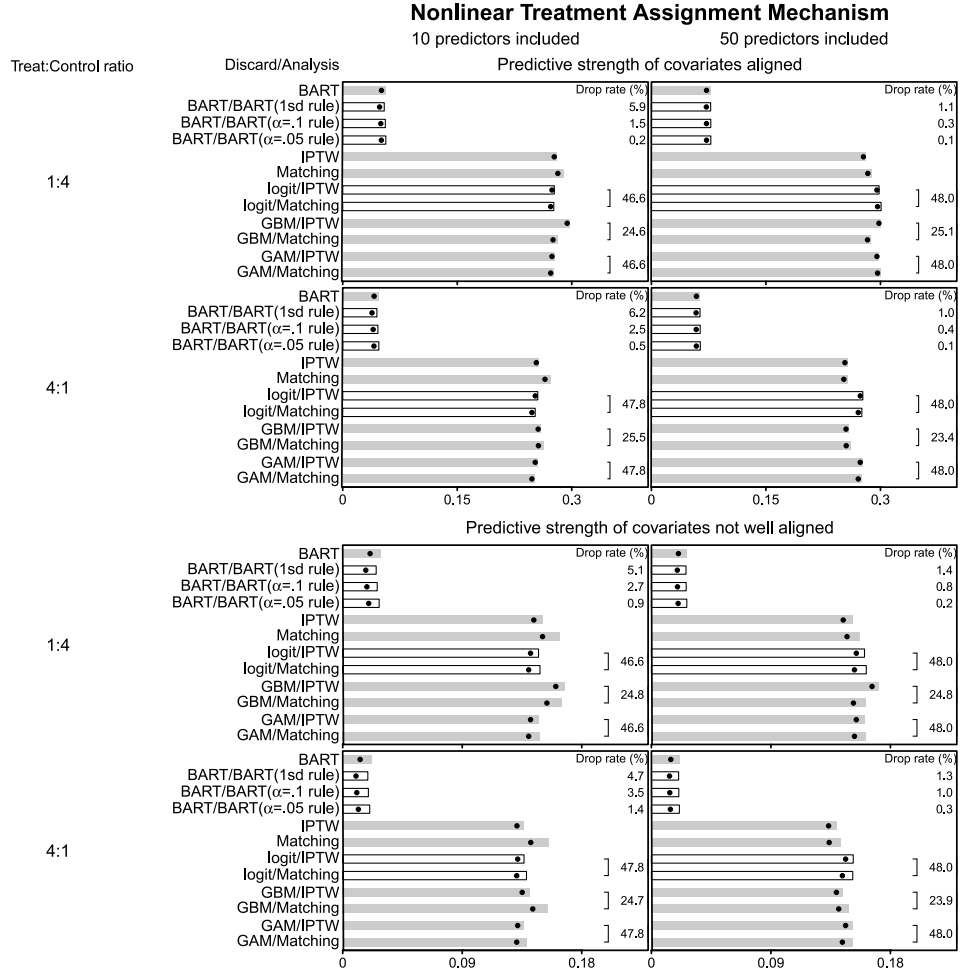}

\caption{Simulation results for the scenarios with nonlinear treatment
assignment and parallel response surfaces. Description otherwise the
same as in Figure \protect\ref{figsimltrpnl}.} \label{figsimnltrpnl}
\end{figure}

The eight plots in Figure \ref{figsimnltrpnl} represent scenarios in
which the
\textit{nonlinear treatment assignment mechanism} was paired with
\textit{parallel
response surfaces}. The nonlinear treatment assignment presents a
challenge to
the naively specified propensity score models. These plots vary between
upper and lower panels in similar ways as seen in Figure \ref
{figsimltrpnl}. Overall, these plots show substantial differences in
results between the BART and propensity score methods.
The BART discard methods drop far fewer observations and yield
substantially less
bias and smaller RMSE across the board. The differences between
propensity score
methods are negligible.

%
\begin{figure}

\includegraphics{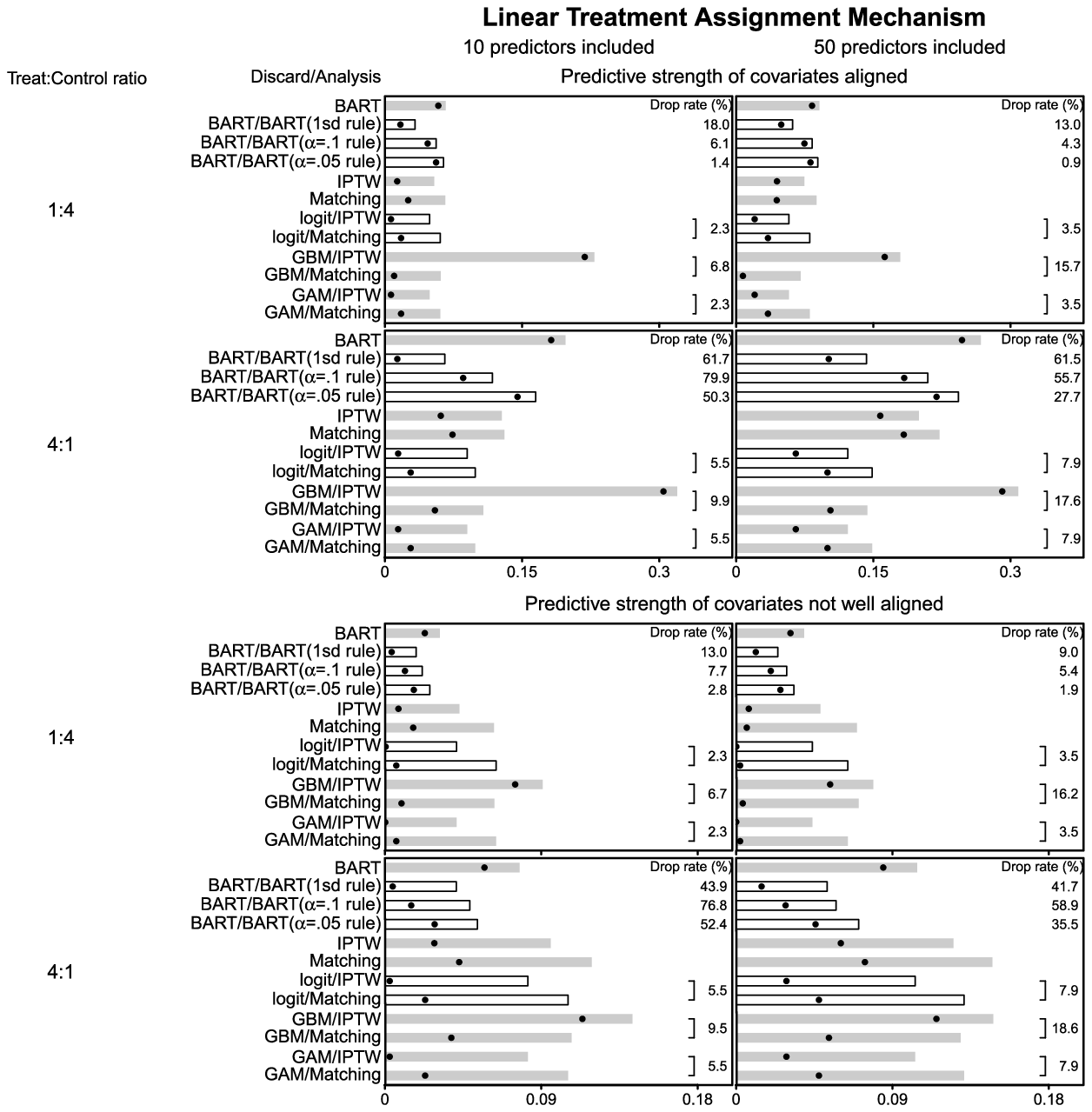}

\caption{Simulation results for the scenarios with nonlinear treatment
assignment and nonparallel response surfaces. Description otherwise
the same as in Figure \protect\ref{figsimltrpnl}.} \label{figsimltrnpnl}
\end{figure}

Figure \ref{figsimltrnpnl} corresponds to scenarios with
\textit{linear treatment
assignment mechanism} and \textit{nonparallel response surfaces}.
The top panel shows little difference in RMSE
or bias for the BART \textit{1 sd rule} compared to the best propensity
score strategies
(sometimes slightly better and sometimes slightly worse). The BART
\textit{$\alpha=0.10$ rule}
and \textit{$\alpha=0.05$ rule} perform slightly worse than the \textit
{1 sd rule} in all four
scenarios.
The bottom panel of Figure \ref{figsimltrnpnl} shows slightly more clear
gains with regard to RMSE for the BART discard methods; the results regarding
bias, however, are slightly more mixed, though the differences are not large.
Across all scenarios the BART \textit{1 sd rule} drops a higher
percentage of treated
observations than the propensity score rules; this difference is
substantial in
the scenarios where treated outnumber controls 4 to 1. The BART
\textit{1 sd rule}
always drops more than the ratio rules when
controls outnumber treated but not when the treated outnumber controls.

%
\begin{figure}

\includegraphics{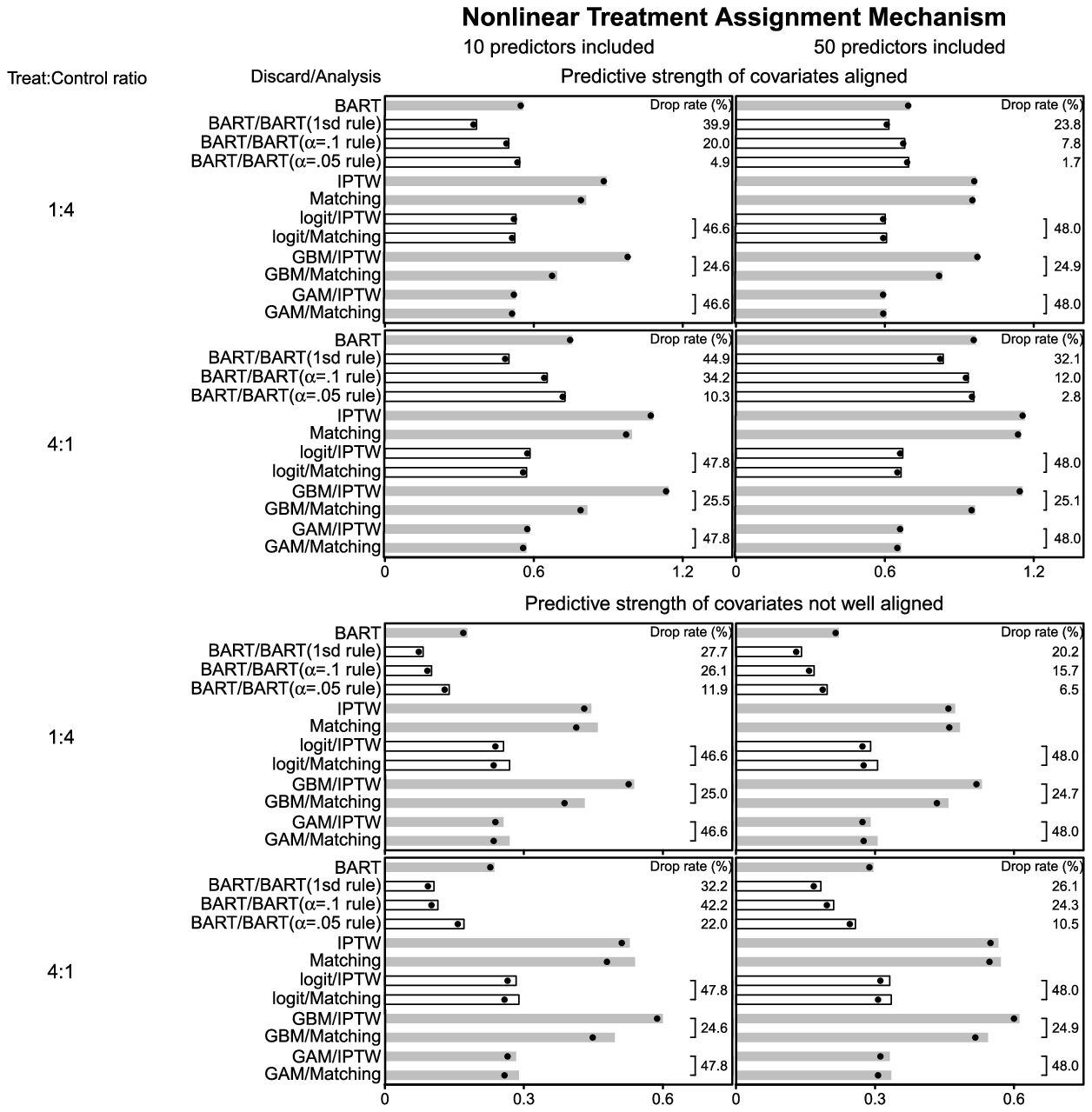}

\caption{Simulation results for the scenarios with nonlinear treatment
assignment and nonparallel response surfaces. Description otherwise
the same as in Figure \protect\ref{figsimltrpnl}.} \label{figsimnltrnpnl}
\end{figure}

The eight plots in Figure \ref{figsimnltrnpnl} all represent
scenarios with
\textit{nonlinear treatment assignment mechanism} and \textit
{nonparallel response
surfaces}. In the top panel the differences between the BART methods and
the best propensity score methods are not large with regard to either bias
or RMSE with BART performing worst in the scenario with 50 potential
predictors and more treated than controls. In the bottom plots corresponding
to misaligned strength of coefficients BART displays consistent gains
over the propensity scores approaches both in terms of bias and RMSE.
All the methods discard a relatively high percentage of treated observations.

While it does not dominate at every combination of our design factors,
the BART
\textit{1 sd rule} appears to perform most reliably across all the methods
overall. In
particular, it almost always performs better with regard to RMSE and it
often performs
well with respect to bias as well.

\section{Discarding and profiling when examining the effect of
breastfeeding on intelligence}
\label{secRealExample}
\begin{figure}

\includegraphics{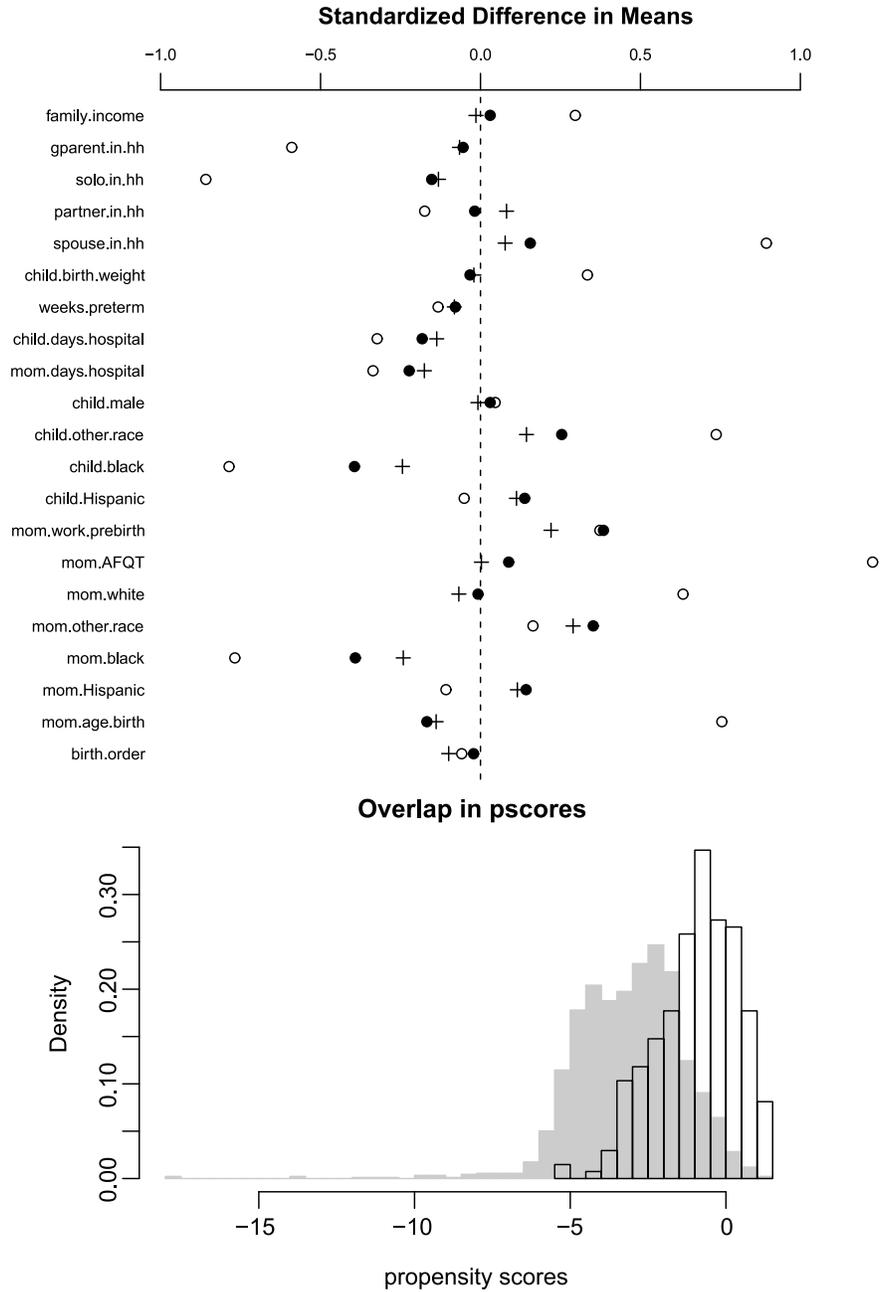}

\caption{Top panel: balance represented as standardized differences in
means for each of three samples: unmatched (open circles),
post-discarding matched (solid circles), and post-discarding
re-weighted (plus signs). Discarding combined with matching and
weighting substantially improve the balance. Bottom panel: overlapping
histograms of propensity scores (on the linear scale)
for both breastfeeding groups.} \label{figoverlap}
\end{figure}

The putative effect of breastfeeding on intelligence or cognitive
achievement has been
heavily debated over the past few decades. This debate is complicated
by the fact that
this question does not lend itself to direct experimentation and, thus,
the vast majority
of the research that has been performed has relied on observational
data. While many
of these studies demonstrate small to medium-sized positive effects
[see, e.g., \citet{andeetal1999,mortetal2002,lawletal2006},
among others] some
contrary evidence exists [notably \citet
{dranloge2000,jainetal2002,deretal2006}].
It has been hypothesized that the effects of breastfeeding increase
with the length of
exposure, therefore, to maximize the chance of detecting an effect, it
makes sense to
examine the effect of breastfeeding for extended durations versus not
at all.
This approach is complicated by the fact that mothers
who breastfeed for longer periods of time tend to have substantially different
characteristics on average than those who never breastfeed (as an
example see
the unmatched differences in means in Figure \ref{figoverlap}).
Thus, identification of areas of common support should be an important
characteristic
of any analysis attempting to identify such effects.

%

Randomized experiments have been performed that address related questions.
Such studies have been used to establish a causal link, for instance,
between two
fatty acids found in breast milk (docosahexaenoic acid and arachidonic
acid) and
eyesight and motor development [see, e.g., \citet{lundetal2010}];
this could represent a piece of the causal pathway between
breastfeeding and subsequent cognitive development. Furthermore, a recent
large-scale study [\citet{krametal2008}] randomized
\textit{encouragement} to
breastfeed and found significant, positive estimates of the
intention-to-treat effect
(i.e., the effect of the randomized encouragement) on verbal and
performance IQ
measures at six and a half years old. Even a randomized study such as
this, however,
cannot directly address the effects of prolonged breastfeeding on
cognitive outcomes.
This estimation would still require comparisons between groups that are
not randomly
assigned. Moreover, an instrumental variables approach would not
necessarily solve
the problem either. Binary instruments cannot be used to identify
effects at different
dosage levels of a treatment without further assumptions. However,
dichotomization
of breastfeeding duration would almost certainly lead to a violation of
the exclusion
restriction.

We examine the effect of breastfeeding for 9 months or more (compared
to not
breastfeeding at all) on child math and reading achievement scores at
age 5 or 6.
Our ``treatment'' group consists of 271 mothers who breastfed at least
38 weeks
and our ``control'' group consists of 1832 mothers who reported 0 weeks of
breastfeeding. To create a cleaner comparison, we remove from our analysis
sample mothers who breastfed greater than 0 weeks or less than 38 weeks.
Given that the most salient policy question
is whether new mothers should be (more strongly) encouraged to breastfeed
their infants, the estimand of interest is the effect of the treatment
on the
controls. That is, we would like to know what would have happened to the
mothers in the sample who were observed to not breastfeed their children
if they had instead breastfed for at least 9 months.

We used data from the National Longitudinal Survey of Youth (NLSY) Child
Supplement [for more information see \citet{chasmottetal1991}]. The
NLSY is a longitudinal survey that began in 1979 with a cohort of approximately
12,600 young men and women aged 14 to 21 and continued annually until
1994 and biannually thereafter. The NLSY started collecting information
on the
children of female respondents in 1986. Our sample comprises 2103 children
of the NLSY born from 1982 to 1993 who had been tested in reading and math
at age 5 or 6 by the year 2000 and whose mothers fell into our two breastfeeding
categories (no months or 9 plus months).

In addition to information on number of weeks each mother breastfed her
child, we also have access to detailed information on potential confounders.
The covariates included are similar to those used in other studies on
breastfeeding
using the NLSY [see, e.g., \citet{deretal2006}], however, we excluded
several post-treatment variables that
are often used, such as child care and home environment measures
since these could bias causal estimates [\citet{rose1984}].
Measurements regarding the child at birth include birth order, race/ethnicity,
sex, days in hospital, weeks preterm, and birth weight. Measurements on the
mother include her age at the time of birth, race/ethnicity, Armed Forces
Qualification Test (AFQT) score, whether she worked before the child
was born,
days in hospital after birth, and educational level at birth. Household measures
include income (at birth), whether a spouse or partner was present at
the time
of the birth of the child, and whether grandparents were present one
year before birth.

The children in the NLSY subsample were tested on a variety of
cognitive measures
at each survey point (every two years starting with age 3 or 4). We
make use of
the Peabody Individual Achievement Test (PIAT) math and reading scores from
assessments that took place either at age 5 or 6 (depending on the
timing of the
survey relative to the age of the child).

To allow focus on issues of common support and causal inference and to
avoid debate about the best way to deal with the missing data,
we simply limit our sample to complete cases. Due to this restriction, this
sample should not be considered to be representative of all children in the
NLSY child sample whose mothers fell into the categories defined.

Comparing the two groups based on the baseline characteristics reveals
imbalance. Figure \ref{figoverlap} displays the balance for the unmatched
(open circles), post-discarding matched (solid circles), and post-discarding
re-weighted (plus signs) samples. The matched and reweighted samples
are much more closely balanced than the unmatched sample, particularly
for the household and race variables.

The bottom panel of Figure \ref{figoverlap} displays the overlap
in propensity scores estimated by logistic regression (displayed on
the linear scale). The histogram for the control units has been shaded
in with grey, while the histogram for the treated units is simply outlined
in black. This plot suggests lack of common support for the control units
with respect to the estimated propensity score. The question remains,
however, whether sufficient common support on \textit{relevant}
covariates exists.

We use both propensity score and BART approaches to address this
question. The results of our analyses are summarized in Table \ref
{tabresultsbf} which displays for each method and test score (reading
or math) combination: treatment effect estimate, standard
error,\footnote{We calculate standard errors for the propensity score
analyses by treating the weights (for matching the weights are equal to
the number of times each observation is used in the analysis) as survey
weights. This was implemented using the \texttt{survey} package in
\texttt{R}.
Technically speaking, uncertainty of each BART estimates is expressed by the
standard deviation of the posterior distribution
of the treatment effect.} and number of units discarded. Without discarding there is
a substantial degree of heterogeneity between BART, linear regression
after one-to-one nearest neighbor propensity score matching with
replacement (Match), IPTW (propensity scores estimated in all cases
using logistic regression), regression and standard linear regression. For reading test
scores the treatment effect estimates are (3.5, 2.5, 1.5, and 3.2) with
standard errors ranging between roughly 0.9 and 1.6. For math test
scores the estimates are (2.4, 3.4, 2.6, and 2.2) with standard errors
ranging between roughly 0.9 and 1.9.

%
\begin{table}
\caption{Table displays treatment effect estimates, associated
standard errors,
and number of units discarded for each method and test score (reading
or math)
combination} \label{tabresultsbf}
\begin{tabular*}{\tablewidth}{@{\extracolsep{\fill}}lc c d{3.0} c c d{3.0}@{}}
\hline
&\multicolumn{3}{c}{\textbf{Reading}} & \multicolumn{3}{c@{}}{\textbf{Math}}\\[-4pt]
& \multicolumn{3}{c}{\hrulefill} & \multicolumn{3}{c@{}}{\hrulefill}\\
& \textbf{Treatment} & \textbf{Standard} & \multicolumn{1}{c}{\textbf{Number}}
& \textbf{Treatment} & \textbf{Standard} & \multicolumn{1}{c@{}}{\textbf{Number}}\\
\textbf{Method} & \textbf{effect} & \textbf{error} & \multicolumn{1}{c}{\textbf{discarded}}
& \textbf{effect} & \textbf{error} & \multicolumn{1}{c@{}}{\textbf{discarded}}\\
\hline
BART & 3.5 & 1.07 & 0 & 2.4 & 1.05 & 0 \\
BART-D1 & 3.5 & 1.07 & 0 & 2.4 & 1.05 & 0 \\
BART-D2 & 3.5 & 1.04 & 93 & 2.4 & 1.04 & 53 \\
BART-D3 & 3.5 & 1.07 & 0 & 2.4 & 1.05 & 0 \\ [4pt]
Match & 2.5 & 1.62 & 0 & 3.4 & 1.74 & 0 \\
Match-D & 3.6 & 1.50 & 168 & 1.5 & 1.13 & 168\\
Match-D-RE & 3.8 & 1.43 & 168 & 1.5 & 1.18 & 168\\
[4pt]
IPTW & 1.5 & 1.57 & 0 & 2.6 & 1.92 & 0 \\
IPTW-D & 1.6 & 1.52 & 168 & 2.6 & 1.85 & 168 \\
IPTW-D-RE & 1.6 & 1.51 & 168 & 2.6 & 1.80 & 168 \\
[4pt]
OLS & 3.2 & 0.87 & 0 & 2.2 & 0.89 & 0 \\
\hline
\end{tabular*}
\end{table}

For the analysis of the effect on reading, the BART \textit{$\alpha=0.10$
rule} would
discard 93
observations, however, neither the BART \textit{1 sd rule} or the \textit
{$\alpha=0.05$ rule} would
discard any.
Regardless of the discard strategy, however, the BART estimate is about 3.5
with posterior standard deviation of a little over 1. Levels of discarding
are similar for math test scores, although for this outcome the BART
\textit{$\alpha=0.10$ rule}
would discard~53. Similarly, the effect estimates (2.4) and associated
uncertainty estimates (a little over 1) are almost identical across strategies.

Using propensity scores (estimated using a logistic model linear in the
covariates) to identify common support discards 168 of the control units.
This strategy does not change depending on the outcome variable. Using
propensity scores estimated on the
remaining units, matching (followed by regression adjustment; Match-D-RE)
and IPTW regression (IPTW-D-RE) yield reading treatment effect
estimates for
the reduced sample of 3.8 (s.e.~1.43) and 1.6 (s.e.~1.51), respectively.
If we do not re-estimate the propensity score after discarding, these estimates
(Match-D and IPTW-D) are 3.6 (s.e. 1.50) and 1.6 (s.e. 1.52),
respectively. The results for
math are quite heterogeneous as well, with matching and IPTW yielding
estimates of 1.5 (s.e. 1.18) and 2.6 (s.e. 1.80), respectively.
Re-estimating the
propensity scores did not change the results for this outcome (when rounding
to the first decimal place).

It is important to remember that the methods that discard units are estimating
different estimands than those that do not, therefore, direct
comparisons between
the BART and propensity score estimates are not particularly informative.
Importantly, however, both
propensity score methods are estimating the same effect (they discarded the
exact same units), therefore, the differences between these estimates
are a bit
disconcerting. One possible explanation for these discrepancies is that
the two
propensity score methods do yield somewhat different results with
regard to
balance as displayed in Figure \ref{figoverlap}; IPTW yields slightly closer
balance on average (though not for every covariate).

What might account for the differences in which units were discarded between
the BART and propensity score approaches?
To better understand, we more closely examine which variables each strategy
identifies as being important with regard to common support by considering
the predictive strength of each covariate with regard to both
propensity score
and BART models in combination with fitting regression trees with the discard
statistics as response variables just as in Section~\ref{ssecprofiling}.

\begin{figure}

\includegraphics{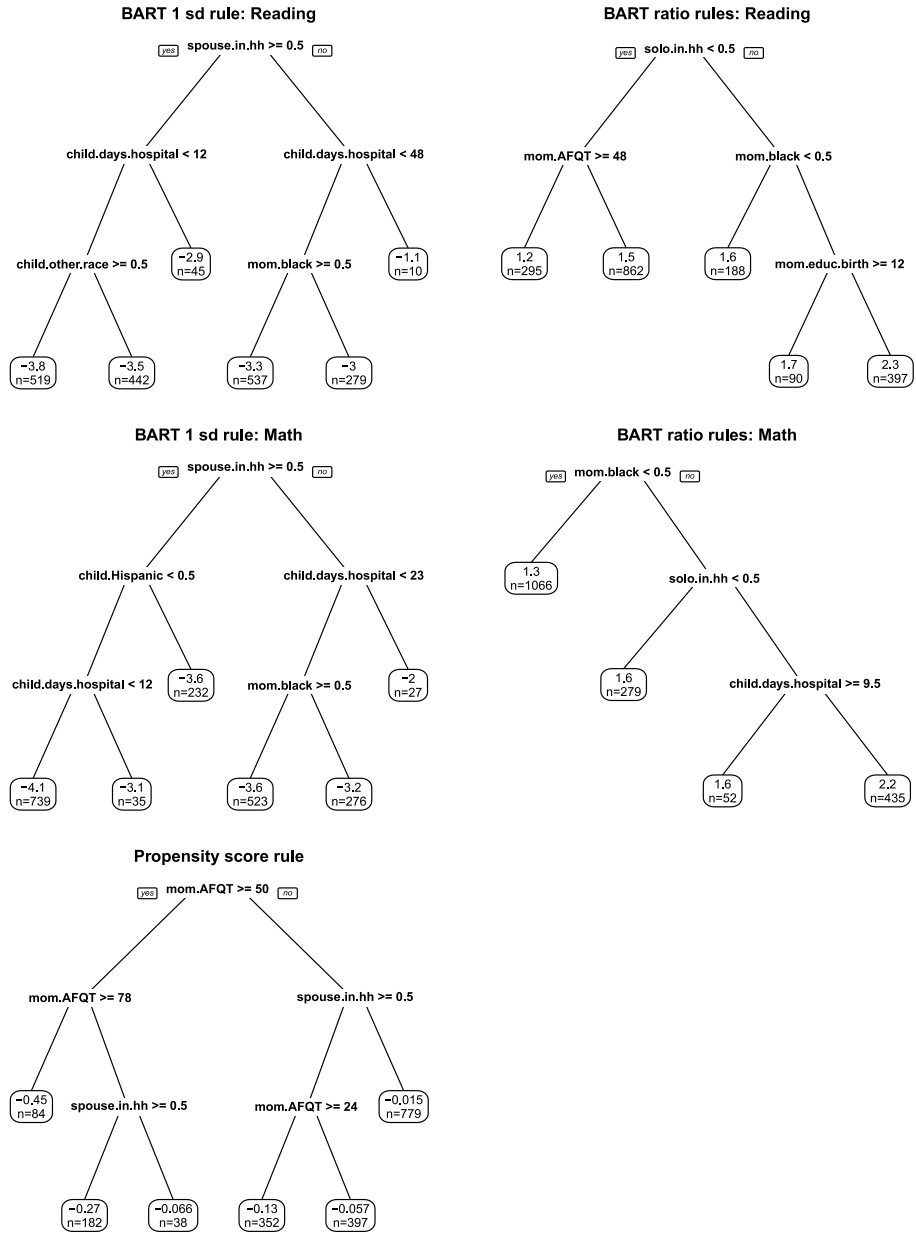}

\caption{Regression trees explore the characteristics of units at risk
of failing to satisfy
common (causal) support. The top two trees use the two statistics from
the BART discard
rules for the reading outcome variable as the response; the next two
trees use the two
statistics from the BART discard rules for the math outcome variable.
The bottom tree uses the estimated propensity score subtracted from
the cutoff (maximum estimated propensity score for the controls). The
predictors of the trees are all the potential
confounding covariates. For all trees the larger the statistic the more
likely the unit
will be discarded, so focus is on the rightmost part of each tree.}
\label{figtreesbf}
\end{figure}

BART identifies birth order, mother's AFQT score, household income,
mother's educational attainment at time of birth, and the number of
days the
child spent in the hospital as the most important continuous predictors for
both outcomes (although the relative importance of each changes a bit
between outcomes). Recall, however, that the BART discard rules
are driven by circumstances in which the level of information about the
outcome changes drastically across observations in different treatment groups.
The overlap across treatment groups for most of these variables is
actually quite
good. While some, like AFQT, are quite \textit{imbalanced}, overlap
still exists
for all of the inferential (control) observations. More problematic in
terms of common
support is the variable that reflects the number of days the child
spent in the hospital;
30 children of mothers who did not breastfeed had values for this
variable higher
than the maximum value (30 days) for the children of mothers who did
breastfeed for nine
or more months. Not surprisingly, this variable is the primary driving
force behind
the BART \textit{1 sd rule} as seen in Figure \ref{figtreesbf}, particularly
for mothers who did not
have a spouse living in the household at the time of birth. Mother's
education plays
a more important role for the BART ratio rules for the reading outcome.
This variable
also has some issues with incomplete overlap and it is slightly more
important in
predicting reading outcomes than math outcomes.

%

A look at the fitted propensity score model, on the other hand, reveals that
breastfeeding for nine or more months is predicted most strongly by the mother's
AFQT scores, her educational attainment, and her age at the time of the
birth of
her child. Thus, these variables drive the discard rule. In particular,
the critical role of mom's AFQT is evidenced in the regression tree for
the discard
rule at the bottom of Figure \ref{figtreesbf}. Children whose
mothers were not married at
birth and whose AFQT scores were less than 50 were most likely to be discarded
from the group of nonbreastfeeding mothers about whom we would like to make
inferences.

What conclusions can we draw from this example? Substantively, if we feel
confident about the ignorability assumption, the BART results suggest a
moderate positive impact of breastfeeding 9 or more months on both reading
and math outcomes at age 5 or 6. The propensity score results for the sample
that remain after discarding for common support are more mixed, with only
the matching estimates on reading outcomes showing up as positive and
statistically significant.

Methodologically, this is an example in which propensity score rules
yield more
discards than BART rules. The most reliable rule based on our
simulation results  (the BART \textit{1 sd rule})
would not discard any units. A closer look at the overlap for specific
covariates
and at regression trees for the discard statistics indicates that
the BART discard rules may represent a better reflection of the actual
relationships
between the variables. The lack of stability of the propensity score
estimates is
also cause for concern. We emphasize, however, that we have used rather naive
propensity score approaches which are not intended to represent best practice.
Given the current lack of guidance with regard to optimal choices for propensity
score models and specific matching and weighting methods, we chose
instead to
use implementations that were as straightforward as the BART approach.

\section{Discussion}\label{secDiscussion}
Evaluation of empirical evidence for the common support assumption has
been given
short shrift in the causal inference literature although the
implications can be important.
Failure to detect areas that lack common causal support can lead to
biased inference
due to imbalance or inappropriate model extrapolation. On the other extreme,
overly conservative assessment of neighborhoods or units that seem to
lack common support may be equally problematic.

This paper distinguishes between the concepts of common support and
common causal
support. It introduces a new approach for identifying common causal
support that
relies on Bayesian Additive Regression Trees\break (BART). We believe that
this method's
flexible functional form and
its ability to take advantage of information in the response surface
allows it to better target
areas of common causal support than traditional propensity-score-based
methods. We
also propose a simple approach to profiling discarded units based on
regression trees.
The potential usefulness of these strategies has been demonstrated
through examples
and simulation evidence and the approach has been illustrated in a real example.

While this paper provides some evidence that BART may outperform
propensity score
methods in the situations tested, we do not claim that it is uniformly
superior or that it
is the only strategy for incorporating information about the outcome
variable. We
acknowledge that there are many ways of using propensity scores that we
did not test,
however, our focus was on examination of methods that were
straightforward to implement
and do not require complicated interplay between the researcher's
substantive knowledge
and the choice of how to implement (what propensity score model to fit,
which matching
or weighting method to use, which variables to privilege in balancing,
which balance
statistics to use). We hope that this paper is a starting point for
further explorations into
better approaches for identifying common support, investigating the
role of the
outcome variable in causal inference methods, and development of more effective
ways of profiling units that we deem to lack common causal support.

There is a connection between this work and that of others [e.g.,
\citet{brooetal2006}] who have pointed out the danger of
strategies that implicitly assign greater importance to variables that
most strongly influence the treatment variable but that may have little
or no direct association with the outcome variable. In response, some
authors such as \citet{kelc2011} have outlined approaches to choosing
confounders in ways that make use of the observed association between
the possible confounders and the potential outcomes. Another option
that is close in spirit to the propensity score techniques but makes
use of outcome data (at least in the control group) would be a
prognostic score approach [\citet{hans2008}]. To date, there has been
no formal discussion of use of prognostic cores for this purpose, but
this might be a useful avenue for further research.\footnote{Thanks to
an anonymous referee for pointing out this connection.}

\section*{Acknowledgments}

The authors would like to thank two anonymous referees and our
Associate Editor, Susan Paddock, for their helpful comments and
suggestions.


%

\printaddresses

\end{document}